\documentclass{article}
\usepackage{spconf,amsmath,graphicx}
\usepackage{cite}
\usepackage{CJK}
\usepackage{multirow}
\usepackage{multicol}
\usepackage[table,xcdraw]{xcolor}
\usepackage{booktabs}
\usepackage{url}
\usepackage{tabularx}
\usepackage[symbol]{footmisc}

\title{AUDIO-VISUAL WAKE WORD SPOTTING SYSTEM FOR MISP CHALLENGE 2021}
\name{
\begin{tabular}{@{}c@{}}
Yanguang Xu$^{\star}$ \qquad 
Jianwei Sun\footnotemark[1]$^{\star}$ \qquad 
Yang Han\qquad
Shuaijiang Zhao\qquad  
Chaoyang Mei \\
Tingwei Guo\qquad
Shuran Zhou\qquad
Chuandong Xie\qquad
Wei Zou\qquad
Xiangang Li\qquad
\end{tabular}}
\address{Beike, Beijing, China \\
\{xuyanguang001, sunjianwei006, zouwei026, lixiangang002\}@ke.com}
\begin{document}
\maketitle

%
%
%

%
\renewcommand*{\thefootnote}{\fnsymbol{footnote}}

\begin{abstract}
This\footnote[0]{$^{\star}$These two authors contributed equally.} paper presents the details of our system designed for the Task 1 of Multimodal Information Based Speech Processing (MISP) Challenge 2021.
The purpose of Task 1 is to leverage both audio and video information to improve the environmental robustness of far-field wake word spotting. 
In the proposed system, firstly, we take advantage of speech enhancement algorithms such as beamforming and weighted prediction error (WPE) to address the multi-microphone conversational audio.
Secondly, several data augmentation techniques are applied to simulate a more realistic far-field scenario.
For the video information, the provided region of interest (ROI) is used to obtain visual representation. 
Then the multi-layer CNN is proposed to learn audio and visual representations, and these representations are fed into our two-branch attention-based network which can be employed for fusion, such as transformer and conformer.
The focal loss is used to fine-tune the model and improve the performance significantly.
Finally, multiple trained models are integrated by casting vote to achieve our final 0.091 score.
\end{abstract}


\renewcommand{\thefootnote}{\arabic{footnote}}
\section{Introduction}
\label{sec:intro}

With the emergence of artificial intelligence (AI) technology, voice controlled applications become widely used in our daily life. A trustworthy wake word spotting (WWS) performance plays a key role in providing a good user experience. A lot of works on audio WWS have been proposed \cite{xie2019target, can2011lattice, kim2019temporal, chen2019small, shan2018attention}. However, in real scenarios, the voices of a wake word may be mixed with the adverse acoustic environments such as background noises (cheers, TV or screams), reverberations, and conversational multi-speaker interactions with a large portion speech overlaps. Eventually, it is difficult for devices to catch the correct instructions only depending on audio. So audio-only wake word spotting methods becoming increasingly challenging in complex scenarios.

Inspired by the multi-modal perception of humans, which is more competent to identify noisy speech, the Audio-Visual (AV) multi-modal has been applied widely in speech community \cite{ma2021end,wu21e_interspeech, zhou21_interspeech,hou2021attention,gao2021visualvoice,rigal21_interspeech,MostafaSadeghi2021SwitchingVA}. The visual information obtained by analyzing lip shapes or facial expressions of the visual modality is more robust than the audio information from complex scenarios. 
Benefitting from deep learning-based algorithms, face recognition\cite{FlorianSchroff2015FaceNetAU} and mouth region tracking\cite{ma2020towards,martinez2020lipreading} become more robust, we can locate ROI precisely, and efficiently extract visual feature representation that is associated with the target talker. 
Researches have been conducted to investigate the possibility to introduce visual information into WWS\cite{liu2014audio,wu2016novel,ding2018audio,momeni2020seeing}.
And how to fuse visual representation with acoustic signal to improve the noise robustness of WWS systems is a core challenge. So, Audio-Visual WWS (AVWWS) is introduced to utilize the visual modality to make up for deficiencies of audio-only WWS in complex acoustic environments.

The first MISP Challenge\footnote{https://mispchallenge.github.io}\cite{chen2022misp} 2021 Task 1 focuses on far-field WWS, which considers the problem of distant multi-microphone conversational audio-visual wake-up and audio-visual speech recognition in everyday home environments. So, how to leverage both audio and video data to improve environmental robustness is a challenging problem.In this paper, we adopt WWS-Transformer and audio-visual deep fusion which two separate encoders learn audio and video modality representation respectively, and then the two modality representations are fused to spot the wake word.

The rest of the paper is organized as follows. Our system description will be shown in Section 2. The experiments and results are reported in Section 3. In the end a summary of the challenge work is given in Section 4.

\section{WWS-Model Unite SYSTEM DESCRIPTION}
\label{sec:format}
This section describes the End-to-End (E2E) AVWWS framework, which consists of the audio-based module, audio-visual-based module, training strategy and system integration. Attention mechanism is introduced into E2E AVWWS framework, A(audio)-Transformer and A-Conformer based on audio information are constructed. Then,considering the loss of audio information, AV-Transformer which is inspired by the Visual Transformer (ViT)\cite{wu2020visual}, based on audio and video information is generated. Finally, the majority voting mechanism is used to complete system integration, and produce the final result.

\subsection{Audio-based module}
In this work, we introduce the A-Transformer which refers to ViT in computer vision shown in Fig.1, the input image patches are replaced with continuous audio frames. Assumed that the input audio features vector $X=\{x_{1},x_{2},...,x_{T}\}$ of $T$ length frames, and position embedding is performed for all audio frames. At first, 512-dimensional vector $C$ initialized by Normal Distribution $N(0, 0.0004)$ as the first frame is added to the output of the Conv-Net which contains two convolution layers and one dense layer, then the spliced vector is sent to multi-layer transformer encoders for attention calculation. Since the convolution layer does not model the information of time dimension, this splicing operation is carried out after the Conv-Net. The first dimension of the sequence vectors, which are generated by the multi-layer transformer encoders, is used as the classification vector. Lastly, the vector is fed to the linear layer with a sigmoid function to output the probability of the wake word. By replacing the transformer encoders with the conformer encoders, we can get the A-Conformer structure.

\begin{figure}
    \centering
    \includegraphics[scale=0.36]{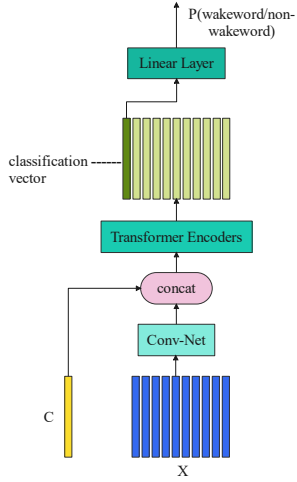}\\
    \caption{A-Transformer.}
    \label{fig: Audio-based module}
\end{figure}

\subsection{Audio-Visual-based module}
Fig. 2 describes two fusion approaches for AV-Transformer. The fusion operation is represented by the red circle. For specific implementation, please refer to formulas (1) and (2).

In the first approach, outputs of A-Conv-Net and V-Conv-Net are fused before sending them to the encoder, as shown in Fig.2(a). In the second approach, two classification vectors generated by A-Transformer encoders and V-Transformer encoders separately will be fused, as shown in Fig.2(b), then the fused vector is used as the input of the linear layer.

\begin{equation}\label{eq:fusion1}
y = w_{a}y_{a} + w_{v}y_{v} 
\end{equation}
\begin{equation}\label{eq:fusion2}
y = y_{a} \cdot y_{v}
\end{equation}

\noindent where $y_{a}$ is the A-Conv-Net's output or the classification vector of A-Transformer encoders, and the corresponding weight value is $w_{a}$. $y_{v}$ is the V-Conv-Net's output or the classification vector of V-Transformer encoders, and the corresponding weight value is $w_{v}$. $y$ represents the fused output.

In the following experiments, weight values used in audio and visual fusion will be iteratively optimized by the training data. The final value of $w_{a}$ is 0.7, whereas $w_{v}$'s final value is 0.3. The performance obtained for both of the approaches will be described in Section 3.

\begin{figure}
    \centering
    \includegraphics[scale=0.35]{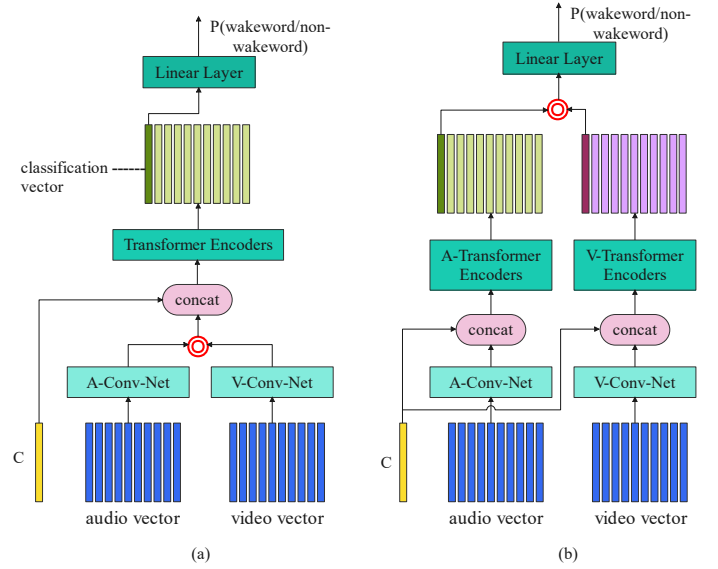}\\
    \caption{AV-Transformer. The red circle is the fusion operation, which represents weighted addition or dot product of vectors.}
    \label{fig:misp_model_architecture3}
\end{figure}

\begin{figure}
    \centering
    \includegraphics[scale=0.4]{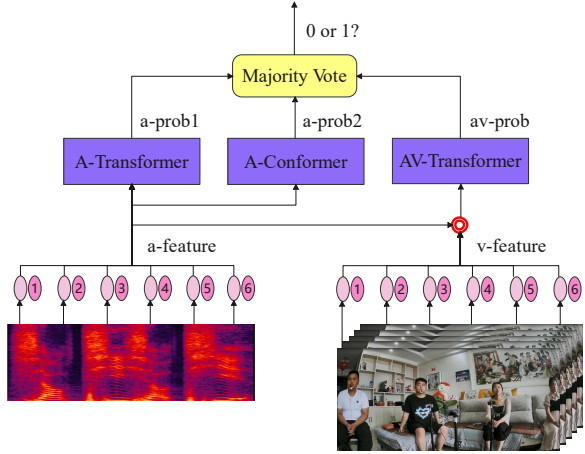}\\
    \caption{WWS-Model Unite System}
    \label{fig:AV-Transformer:}
\end{figure}

\begin{table*}[hbt]
  \caption{Performances of audio-only WWS models.}
  \label{tab:audio_kws}
  \centering
  \scalebox{0.87}{
  \begin{tabular}{llllll}
    \toprule
    Model & \#Encoder &  Augmentation   & Loss  & Dev  & Eval  \\
    \midrule
    \multirow{4}{*}{A-Transformer} & 1  & / & CE & 0.153 & 0.207 \\
    & 4  & / & CE & 0.127 & 0.175 \\
    & 4  & / & CE \& Focal & 0.120 & 0.149 \\
    & 4  & SpecAugment & CE \& Focal & 0.109 & 0.106 \\
    \midrule
    A-Conformer & 4 & SpecAugment & CE \& Focal & 0.105 & 0.116 \\
    \bottomrule
  \end{tabular}
  }
\end{table*}

\subsection{Training strategy}

Compared with the cross-entropy (CE) loss, focal loss can make the training model better mine and learn some difficult or misclassified samples, thereby effectively improving the accuracy of the classification task. In training of our designed system, the WWS tasks are trained with CE and focal loss in two stages respectively. In the first stage, the base model is trained based on CE loss. In the second stage, the loss function is changed into focal CE loss, and all data used in the first stage are used to fine-tune the obtained base model. Finally, the fine-tuned model will be used for decoding. This method can also solve the problem of serious imbalance of positive and negative proportion in training samples.





\subsection{System integration}

The AV-Transformer uses much more information of visual, and should have better performance compared with the A-Transformer theoretically. However, image features of MISP are extracted from the middle-field and far-field videos, especially there are multi-speaker in the far field videos. Therefore, the effective image feature is difficult to be extracted from these videos which can have some bad cases during model training and test. In order to solve this question, we use a majority vote method to integrated these models (A-Transformer, A-Conformer, AV-Transformer) to get the final result, as expressed in Fig. 3 and Formula (3).
\begin{equation}\label{eq:system fusion}
\begin{split}
result_{fusion} = Mo(result_{A-Transformer},\\ result_{A-Conformer},\\
result_{AV-Transformer})
\end{split}
\end{equation}
The score outputted by each model will be compared with the threshold value to determine whether the result is 0 or 1. “Mo” means taking the mode of three models' results.

\label{sec:analysis}

\label{sec:analysis}

\section{Experiments}
\label{sec:pagestyle}

Data preprocessing, feature extraction, model training and decoding are based on Athena\footnote{https://github.com/athena-team/athena.git} and Kaldi\footnote{https://github.com/kaldi-asr/kaldi.git}. Details are described in the following subsections.

\begin{table*}[ht]
  \caption{Performances of AVWWS models. A-Conv and V-Conv denote the weighted outputs of A-Conv-Net and V-Conv-Net mentioned above. A-Attention and V-Attention denote the weighted classification vectors generated by A-Transformer and V-Transformer encoders.}
  \label{tab:audio_kws}
  \centering
  \scalebox{0.87}{
  \begin{tabular}{llllll}
    \toprule
    Model & Fusion &  Augmentation   & Loss  & Dev \\
    \midrule
    \multirow{4}{*}{AV-Transformer} & A-Conv+V-Conv  & / & \multirow{4}{*}{CE} & 0.160 \\
    & A-Conv*V-Conv  & / & & 0.204 \\
    & A-Attention+V-Attention  & / & & 0.132 \\
    & A-Attention*V-Attention  & / & & 0.191 \\
    \bottomrule
  \end{tabular}
  }
\end{table*}

\begin{table*}[ht]
  \caption{Performances of all WWS models in streaming and E2E type}
  \label{tab:all_kws}
  \centering
  \scalebox{0.87}{
  \begin{tabular}{lllllll}
    \toprule
    WWS Type & Model & Model Detail  & Data & Loss & Dev & Eval  \\
    \midrule
    Streaming & CNN-DNN & 2 Conv+3 Dense & 60h pos+200h neg & CE & 0.314 & / \\
    \midrule
    \multirow{7}{*}{E2E} & \multirow{3}{*}{CRNN} & 2 Conv+2 biGRU & 60h pos+200h neg & \multirow{3}{*}{CE} & 0.209 & / \\
    \cline{3-4}\cline{6-7}
    & & \multirow{2}{*}{2 Conv+5 biLSTM} & 60h pos+200h neg & & 0.186 & / \\
    \cline{4-4}\cline{6-7}
    & & & 170h pos+530h neg & & 0.178 & / \\
    \cline{2-7}
    & A-Transformer & \multirow{2}{*}{2 Conv+4 encoders+1 Dense} & \multirow{2}{*}{170h pos+530h neg} & \multirow{2}{*}{CE\&Focal} & 0.109 & 0.106 \\
    \cline{2-2}\cline{6-7}
    & A-Conformer & & & & 0.105 & 0.116 \\
    \cline{2-7}
    & \multirow{2}{*}{AV-Transformer} & 2 Conv+4 AV-encoders & A
(170h pos+530h neg) & \multirow{2}{*}{CE} & \multirow{2}{*}{0.132} & \multirow{2}{*}{/} \\
    & & +1 Dense & V(Far 124h) & & & \\
    \midrule
    \multicolumn{5}{c}{Majority Vote} & / & 0.091 \\
    \bottomrule
  \end{tabular}
  }
\end{table*}

\subsection{Dataset}

The dataset provided by MISP Challenge 2021 Task 1, which includes the same amount of near-field, mid-field and far-field data, consists of 118.53 hours training set, 3.39 hours development set and 2.87 hours evaluation set. In training and development set, there are positive cases of 5.67 hours and 0.62 hours respectively.

Before training, the data are processed by the following steps: 

1) Audio normalization: negative audios in the training set are clipped into approximately the same length with negative audios in the development set.

2) Speech enhancement: we implement a fixed directional beamformer that enhances the 90 degree direction, and followed by WPE dereverberation.

3) Data augmentation: we generate room impulse response by using the pyroomacoustic tool\footnote{https://github.com/LCAV/pyroomacoustics.git}, and convolve them with original near-field data to simulate mid and far-field data. Speed perturbation is used to generate 0.9 and 1.1 speed ratio data. At last, we mix noises provided by the officials with a random SNR from -15 to 15 dB.

4) Feature extraction: 63-dimensional fbank is extracted from audios, and all the features are normalized with global CMVN. SpecAugment\cite{DanielSPark2019SpecAugmentAS} is used for data coverage and model robustness. In addition, video features are extracted by open source tools provided by the MISP organizer. Audio and video features will be unified into the same dimension through two separate Conv-Nets.

Finally, we expand the audio training set to a total of 700 hours, of which 170 hours are positives and 530 hours are negatives. Besides, we only use far-field video data during all the training steps.

\subsection{Model details and Evaluation}

For the transformer or conformer structure, we use 4 self-attention blocks, each with 8 heads, 512-dimensional hidden size, and 2,048 dimensions for feed-forward layer. All models are trained on 1 GPU V100, and the batch size is 32. The learning rate is set to 0.00001 initially and warmed up at the first 10,000 steps by the Adam optimizer. 

The final evaluation score of the model is determined by the false rejection rate(FRR) and the false acceptance rate (FAR), and the calculation is as follows.

\begin{equation}\label{eqFRR}
FRR = \frac{FN}{FN + TP} 
\end{equation}
\begin{equation}\label{eq:FAR}
FAR = \frac{FP}{FP + TN} 
\end{equation}
\begin{equation}\label{eq:score}
SCORE_{WWS} = FRR + FAR 
\end{equation}

\noindent where \textbf{\textit{FN}} denotes the number of cases erroneously divided into negative, \textbf{\textit{TP}} denotes the number of cases correctly divided into positive, \textbf{\textit{FP}} denotes the number of cases erroneously divided into positive, and \textbf{\textit{TN}} denotes the number of cases correctly divided into negative.

\subsection{Results}

Firstly, we explore some E2E model structures with different loss functions and the results are shown in Table 1.
Then focal loss and SpecAugment are applied for audio data, and we get a better of 0.106 by using a single model based on A-Transformer on evaluation set. 
Secondly, multi-mode information is fused to build AV-Transformer to enhance the performance. Table 2 shows the performance of AV-Transformer models which use different fusion methods. 
The performance of CNN-DNN streaming model and various E2E models are listed in Table 3. We unit all kinds of E2E attention models with a majority vote mechanism to get the best result. 

These conclusions can be drawn from our experiments:

(1) The E2E model based on Transformer or Conformer performs better than CRNN model and CNN-DNN streaming model when using audio data only.

(2) Focal loss and SpecAugment can further improve the final performance.

(3) The AV-Transformer is effective to reduce error score by majority vote.

(4)  The majority vote has taken the votes from three top performed models originated from our A-Transformer, A-Conformer and AV-Transformer, which yields the final 0.091 score. 


\section{CONCLUSION}
\label{sec:typestyle}

We propose an AV-WWS-Model Unite System for the task of far-field wake word spotting in complex scenarios. 
The proposed system mainly contains three subsystems: A-Transformer, A-Conformer and AV-Transformer and these models are trained either by cross-entropy loss or focal loss. In order to take advantage of audio and video information, a majority vote is adopted for these models to get the final result. The experiments results prove that our AV-WWS-Model Unite System yields better performance.


\newpage
\bibliographystyle{IEEEbib}
\small
\bibliography{main}
\end{document}